\documentclass{nature}
\usepackage{graphicx}
\usepackage{booktabs}
\usepackage{caption}
\usepackage{amsmath}
\usepackage{amssymb}

\title{Solid-state laser refrigeration of a semiconductor optomechanical resonator}

\author{Anupum Pant$^{1}$, Xiaojing Xia$^2$, E. James Davis$^3$ \& Peter J. Pauzauskie$^{1,4}$}

\begin{document}

\maketitle

\begin{affiliations}
 \item Materials Science and Engineering  Department,  University of Washington, Seattle, Washington 98195
  \item Department of Molecular Science \& Engineering, University of Washington, Seattle, WA, USA
 \item Chemical Engineering  Department,  University of Washington, Seattle, Washington 98195
 \item Physical \& Computational Sciences Directorate, Pacific Northwest National Laboratory, Richland, WA 99352
\end{affiliations}

\begin{abstract}

\end{abstract}

\section*{Introduction}
\textbf{Photothermal heating represents a major constraint that limits the performance
of many nanoscale optoelectronic and optomechanical devices including
nanolasers\cite{li_room-temperature_2017}, quantum optomechanical
resonators\cite{teufel2011sideband, chan2011laser}, and integrated photonic
circuits\cite{garcia-meca_-chip_2017}.  Although radiation-pressure damping has been
reported to cool an individual vibrational mode of an optomechanical resonator to its
quantum ground state\cite{teufel2011sideband, chan2011laser}, to date the internal
material temperature within an optomechanical resonator has not been reported to cool
via laser excitation. Here we demonstrate the direct laser refrigeration of a
semiconductor optomechanical resonator $>$20K below room temperature based on the
emission of upconverted, anti-Stokes photoluminescence of trivalent ytterbium ions doped
within a yttrium-lithium-fluoride (YLF) host crystal. Optically-refrigerating the lattice of a
dielectric resonator has the potential to impact several fields including scanning probe
microscopy\cite{meyer1996simple, kirk1988low, hosseini2014multimode}, the sensing of
weak forces\cite{braginskii1977measurement, abramovici1992ligo}, the measurement of
atomic masses\cite{jensen2008atomic}, and the development of radiation-balanced
solid-state lasers\cite{bowman2005}. In addition, optically refrigerated resonators may
be used in the future as a promising starting point to perform motional cooling for
exploration of quantum effects at mesoscopic length scales\cite{schwab2005putting}, temperature control within integrated photonic devices\cite{garcia-meca_-chip_2017}, and solid-state laser refrigeration of quantum materials\cite{kolkowitz2012coherent}.}

Photothermal heating is a perennial challenge in the development of advanced optical
devices at nanometer length scales given that a material's optical index of refraction, bandgap, and Young's modulus all vary with temperature. For instance, reducing the
mechanical motion of an optomechanical resonator to its quantum ground state requires
that the temperature ($T$) must be much less than $h\nu/k_b$, where $\nu$ is the mode
frequency, $h$ and $k_b$ are Planck and Boltzmann constants,
respectively\cite{o2010quantum}. Critically, incident laser irradiances must be kept low
enough to avoid photothermal heating of the resonator above cryogenic
temperatures\cite{teufel2011sideband, groblacher2009demonstration, park2009resolved,
o2010quantum, chan2011laser}. Here, we demonstrate a new approach for the photothermal
cooling of nanoscale optoelectronic devices through the emission of blue-shifted
(anti-Stokes) photoluminescence.  In particular, we used a micron-scale grain of 10\%
Yb$^{3+}$-doped YLiF$_4$ (Yb:YLF) located at the end of a semiconductor optomechanical resonator (CdS) to cool the resonator $>$20K below room
temperature following excitation with a cw-laser source with wavelength $\lambda_0$ =
1020 nm.

The idea of refrigerating metallic sodium vapors using anti-Stokes luminescence was
first proposed by Pringsheim in 1929\cite{pringsheim1929zwei}.  Following the
development of the laser, Doppler cooling of metallic vapors led to the first
observation of Bose-Einstein condensates in 1995\cite{anderson_observation_1995}. The
first experimental report of solid-state laser cooling came in 1995 using Yb$^{3+}$
doped ZBLANP glass (Yb:ZBLANP) \cite{epstein1995observation}. Since then, two decades of
research in the area of solid-state laser refrigeration has culminated in the
development of a solid-state optical cryo-cooler using bulk Yb:YLF single crystals grown
using the Czochralski method\cite{melgaard2016solid}, which has cooled crystals to 91 K from room
temperature.  The primary advantage of using crystalline materials for
solid-state laser cooling is the existence of well-defined crystal field levels which
minimizes inhomogenous broadening of rare-earth absorption spectra.  Recently this has
enabled the first experimental demonstrations of cold Brownian motion
\cite{roder2015laser} since Einstein's seminal paper\cite{Einstein} on Brownian motion
in 1905. The increased optical entropy of the blue-shifted photons makes this
cooling cycle consistent with the second law of
thermodynamics\cite{landau_thermodynamics_1946}.

Recently it has been claimed that semiconducting cadmium sulfide nanoribbons (CdSNRs) suspended over a silicon wafer can be refrigerated optically through upconverted, anti-Stokes photoluminescence from the CdSNR\cite{zhang2013laser}. In contrast, in this work the semiconductor optomechanical resonator is suspended in vacuum from a silicon wafer  to reduce the potential for photothermal heating of the adjacent silicon substrate. Van der Waals bonding is used to attach a low-cost, hydrothermal ceramic Yb:YLF microcrystal\cite{roder2015laser} to the end of the resonator cavity. Rare-earth (Yb\textsuperscript{3+}) point-defects within the YLF emit anti-Stokes photoluminescence which cools both the YLF microcrystal, and also the underlying semiconductor optomechanical resonator. The YLF serves both as a local thermometer (discussed in more detail below) and also as a heat sink which extracts thermal energy from the cantilever, increasing its Young's modulus, and thereby blue-shifting the cantilever's
optomechanical eigenfrequency. The transmitted laser causes minimal
heating of the cantilever supporting the YLF crystal due to its small thickness (150 nm) and extremely low
absorption coefficient of CdS at 1020 nm\cite{treharne2011optical}. The temperature of the source and the cantilever system were measured using two independent non-contact temperature measurement methods -
differential luminescence thermometry\cite{seletskiy2013precise} and optomechanical
thermometry\cite{pant2018optomechanical}, respectively, which agree well with each other.

\begin{figure}
\label{F1}
\includegraphics[width = 16cm]{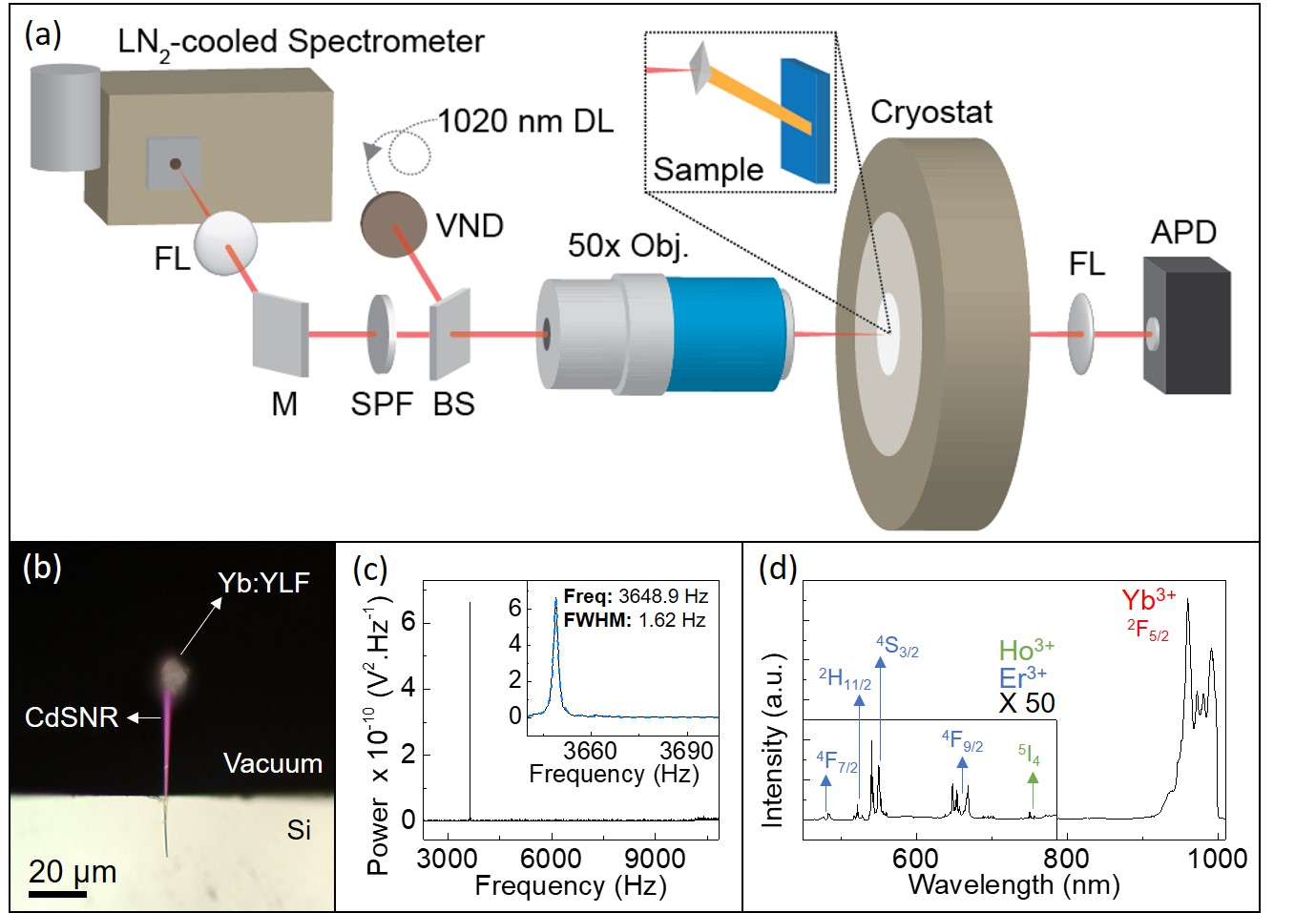}
\caption{a) The schematic of the eigenfrequency and upconverted fluorescence measurement setup. FL, M, SPF, DL, BS, VND and APD stand for focusing lens, mirror, 1000 nm short pass filter, diode laser, beam splitter, variable neutral density filter and avalanche photodiode, respectively.
b) A bright field optical image of the CdSNR cantilever supported using a silicon substrate with a Yb$^{3+}$:YLF crystal placed at the free end.
c) A peak in the thermomechanical noise spectrum originating from the fundamental eigenfrequency of the CdSNR with Yb$^{3+}$:YLF sample obtained at the 0.038 MW/cm$^2$ at 300 K.
d) A stitched, up-converted fluorescence spectrum obtained at room temperature using a 1020 nm excitation source (0.039 MW/cm$^2$) focused on the suspended Yb$^{3+}$:YLF crystal. A 1000 nm short pass filter was used to cut off the laser line.}
\end{figure}

\section*{Optomechanical Thermometry}

A CdSNR was placed at the end of a clean silicon substrate, and a hydrothermally grown
10\% Yb:YLF crystal was placed at the free end of the CdSNR cantilever. CdS was chosen because of its wide band gap and low-cost, though in principle any material with low near-infrared (NIR) absrption can be used.  A bright-field
optical image of a representative sample is shown in Fig. 1b. The silicon substrate
was loaded inside a cryostat chamber such that the free end of the cantilever was
suspended over the axial hole in the cryostat, and the system was pumped to
$\sim$10$^{-4}$ torr. As shown in Fig. 1a, a 1020 nm laser was focused onto the
Yb:YLF crystal at the end of the cantilever. The time-dependent intensity of the
forward-scattered 1020 nm laser was measured by focusing it onto an avalanche photodiode
(APD). To
measure the cantilever's eigenfrequencies the voltage vs. time signal was Fourier-transformed to obtain its thermomechanical noise spectrum\cite{lee2008temperature, pini2011shedding}. A representative power spectrum measured on the sample at
300 K using a laser irradiance of 39 kW/cm$^2$ is shown in Fig. 1c. A sharp peak,
fitted using a standard Lorentzian with a peak position at 3648.9 Hz, was attributed to
the first natural resonant frequency mode (``diving board mode") of the fluoride crystal
on the nanoribbon (FCNR) cantilever system (see Supplementary Information). As shown in Fig. 1a,
the backscattered photoluminescence was collected from the rear end of the objective,
transmitted through a beamsplitter, was filtered using a 1000 nm short-pass filter and focused into the spectrometer slit. A photoluminescence (PL)
spectrum was recorded at different grating positions, with appropriate collection times
to avoid saturating the detector, and were stitched together. Ten spectra were collected using 39 kW/cm$^2$ of laser irradiance for 0.1 s and averaged. The intense Yb$^{3+}$ transitions\cite{bensalah2004growth, rahman2017laser} in the range of 800 to 1000 nm, with major peaks at 960 (E$_6$-E$_1$), 972 (E$_5$-E$_1$) and 993 (E$_5$-E$_3$) nm were observed (Fig. 1d). A longer acquisition time (50x) was used to collect the weaker luminescence signal from the other rare earth (RE) impurities that were not explicitly added during synthesis. The up-converted green and red emission peaks at 520, 550 and
650 nm are attributed to the transitions from  $^2$H$_{11/2}$, $^4$S$_{3/2}$ and $^4$F$_{9/2}$, respectively, of trivalent erbium ions (Er$^{3+}$)\cite{reddy1994infrared,yeh1991intensity}. Other minor transitions are labeled.

\begin{figure}
\includegraphics[width = 16cm]{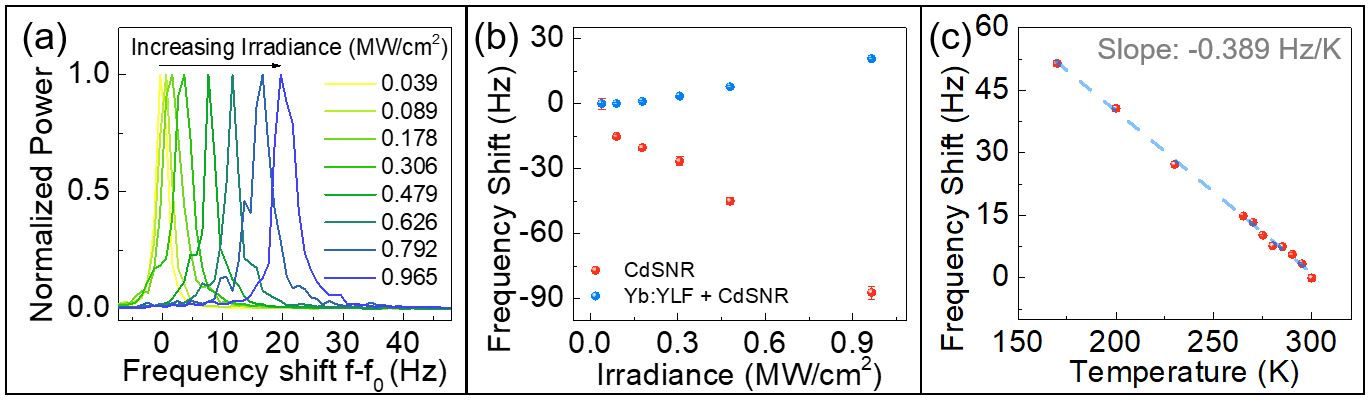}
\caption{a) Normalized power spectra for a representative laser refrigeration measurement at each laser irradiance with an ambient reference temperature of 295 K ($f_0$ = 3632.2 Hz).
b) The frequency shift ($f-f_0$) with laser power at 295 K for both a plain CdSNR (red) and CdSNR with Yb:YLF (blue). Each data point was averaged from Lorentzian fits of 6 power spectra and error bars represent one standard deviation. Note that for small standard deviations, the error bars overlap with the data point. $f_0$ is 3632.2 Hz and 17384.4 Hz, respectively.
c) Temperature calibration of the CdSNR with Yb:YLF obtained by measuring the frequency shift $f-f_0$ ($f_0$ = 3653.6 Hz) as a function of the cryostat temperature.}
\label{F2}
\end{figure}

Power spectra normalized using the maximum value at different laser irradiances obtained from the sample are plotted in Fig. 2a. When fit to a standard Lorentzian, the peak values show a blue-shift in the
eigenfrequency of the FCNR system as the laser power is increased. The fitted peak
values of these power spectra are shown in Fig. 2b. As the laser irradiance was
increased, the Yb:YLF source reached lower temperatures, thereby extracting more heat
from the CdSNR cantilever and causing a blue-shift in the frequency due to an increased
Young's modulus of the CdS at lower temperatures. Using 0.5 MW/cm$^2$ of 980 nm laser resulted in the irreversible photothermal melting of the cantilever device shown in Fig. S12. When the Yb:YLF crystal was removed
from the CdSNR cantilever in Fig. 2b, the fundamental frequency measured at 39 kW/cm$^2$ increased
to a higher value of 17384.3 Hz due to the removal of mass from the system ($\sim$1.3x10$^{-9}$g, see Supplementary Information). As a control experiment, the eigenfrequency of the CdSNR
cantilever itself was measured after the removal of the Yb:YLF crystal. The eigenfrequency of the cantilever without the crystal was then measured as a function of the laser power and is shown in Fig. 2b. The eigenfrequency red-shifts as the laser irradiance is increased,
suggesting greater heating of the cantilever at higher irradiances due to the decreasing Young's modulus at higher temperatures\cite{pant2018optomechanical}. The temperature
of the FCNR device was calibrated by increasing the temperature of the cryostat
from 160 to 300 K, which showed a linear red-shift in the eigenfrequency of the
cantilever. The slope of -0.389 Hz/K obtained using this calibration was used to measure
the temperature change of the cantilever system during laser refrigeration experiments.
The maximum blue-shift in the eigenfrequency as a function of laser irradiance of the
was +20.6 Hz at an irradiance of 965 kW/cm$^2$, compared to the lowest irradiance of 39
kW/cm$^2$. Based on the isothermal temperature calibration, ignoring temperature
gradients and other optomechanical effects on the cantilever due to increased
irradiance, this blue-shift of +20.6 Hz corresponds to a temperature change of 53 K
below room temperature (assuming a negligible change in temperature at a laser
irradiance of 39 kW/cm$^2$).

However, to obtain the absolute change in temperature it is important to view the system
through a modified Euler-Bernoulli beam theory and include the effects of the laser
trapping forces\cite{ashkin1997optical} on the Yb:YLF crystal, which acts as a spring at the end of the
cantilever. With increased irradiances, due to the increased force constant of the
spring, the eigenfrequency of the cantilever increases. Analytically the eigenfrequency
($f_i$) in hertz, of a uniform rectangular beam is given
by\cite{balachandran2008vibrations}:
\begin{equation}
    f_i = \frac{1}{2\pi}\frac{\Omega_i^2}{L^2}\sqrt{\frac{EI}{\rho}}.
\end{equation}
Here $L$ is the length, $\rho$ is the linear density, $E$ is the Young's modulus, and
$I$ is the area moment inertia of the cross section of the beam. The $i^{th}$ eigenvalue
of the non-dimensional frequency coefficient $\Omega_i$ satisfies the following equation
for a uniform rectangular cantilever with a mass $M_0$ and spring of spring constant $K$
attached at the free-end of the cantilever of mass $m_0$.
\begin{equation}
    -\Bigg(\frac{K}{\Omega_i^3}-\frac{\Omega_i M_0}{m_0}\Bigg)[\cos(\Omega_i)\sinh(\Omega_i)-\sin(\Omega_i)\cosh(\Omega_i)]+\cos(\Omega_i)\cosh(\Omega_i)+1 = 0.
\end{equation}
To experimentally probe the effects of the laser trapping forces, the power-dependent
eigenfrequency measurements were performed at a constant cryostat temperature of 77 K.
At temperatures as low as 77 K, the cooling efficiency of the Yb:YLF crystal decreases
due to diminishing resonant absorption and red-shifting of the mean fluorescence
wavelength\cite{seletskiy2013precise}. Due to negligible cooling with increased
irradiance, and with the equilibrium temperature being maintained by the crysostat, it
is assumed that any blue-shift in the eigenfrequency of the system was solely due to
the greater laser trapping force at higher irradiance. Therefore, the excessive
blue-shift at room temperature (6 +/- 2.2 Hz ) can be attributed to the change in Young's modulus
due to cooling of the CdSNR cantilever. According to this calibration, the cantilever's
temperature is reduced 15.4 +/- 5.6 K below room temperature. Since cantilever eigenfrequencies are calibrated at isothermal conditions, the temperatures measured via cantilever eigenfrequencies during laser-refrigeration do not directly measure the coldest point within the cantilever, but rather a lower bound of the absolute minimum achievable temperature decrease\cite{pant2018optomechanical}. This is a consequence of temperature gradients within the cantilever that lead to gradients of the cantilever's Young's modulus. Based on finite element eigenfrequency modeling of the cantilever with a spatially varying Young's modulus, the coldest point in the cantilever can be calculated (see Supplementary Information). Below we present a steady-state, heat-transfer model of the laser-cooled cantilever system to quantify how thermal gradients within CdSNR cantilevers affect eigenfrequency measurements during laser cooling experiments.

\section*{Heat Transfer Analysis}

\begin{figure}
\includegraphics[width = 16cm]{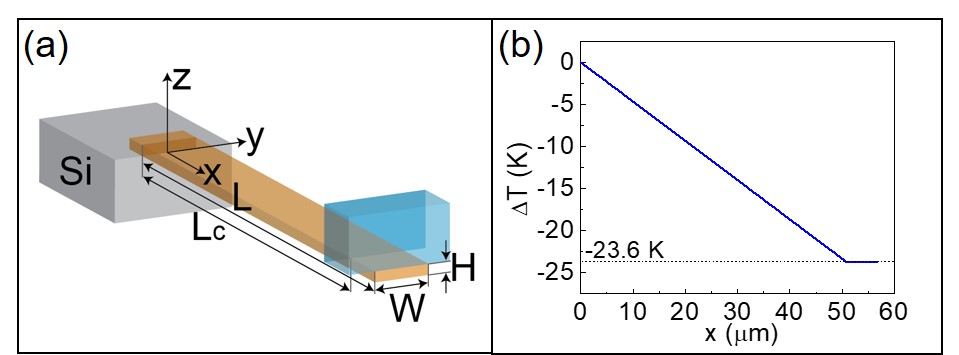}
\caption{a) The geometry of FCNR system used for analytical and finite element heat transfer modeling. b) The steady state temperature along the length of the CdSNR calculated using analytical one dimensional solution obtained assuming all of the cooling power produced by the YLF crystal flows through the CdSNR cross section at L$_c$.}
\label{F3}
\end{figure}

A cantilever of length `$L$', width `$W$', and thickness `$H$' is modeled with a YLF crystal placed at the free end (see Fig. \ref{F3}a).   The YLF crystal is approximated as a cuboid with sides of $H_c$ = 6, $L_c$ = 7.5 and $W_c$ = 6 $\mu$m, such that the volume and aspect ratio was similar to the tetragonal bi-pyramidal YLF crystal used experimentally.

At steady state the temperature distribution in the nanoribbon satisfies the energy equation given by:
\begin{equation}
\frac{\partial^2 T}{\partial x^2}+\frac{\partial^2 T}{\partial y^2}+\frac{\partial^2 T}{\partial z^2} = 0.
\label{eq:pde}
\end{equation}
Heat transfer to the surroundings by conduction and convection is absent due to the vacuum surrounding the cantilever.  Radiant (blackbody) energy transfer to or from the surroundings is negligible due to the relatively low temperatures of the cantilever and its small surface area.  Therefore, the heat flow within the cantilever is one-dimensional and equation (\ref{eq:pde}) reduces to:
\begin{equation}
\frac{d^2 T}{d x^2} = 0,
\label{eq:pde_2}
\end{equation}
which has the general solution:
\begin{equation}
T(x) = C_1x+C_2.
\label{eq:pde_gen.sol}
\end{equation}

Assuming negligible interfacial resistance between the cantilever and the underlying silicon substrate, the temperature at the silicon/CdS interface
(x = 0) is the cryostat temperature $T_0$. Consequently, the boundary condition at the
base of the nanoribbon is $T(0)=T_0$. If all of the heat generated in the YLF crystal
is transferred to or from the CdSNR across the interface at $x = L_c$, the heat flux at
the interface is given by:
\begin{equation}
\kappa\frac{dT}{dx}(L_c) = \frac{\dot{Q}_{c}}{HW},
\end{equation}
in which $\dot{Q}_{c}$ is the rate of heat removal from the YLF crystal, and $\kappa$ is the thermal conductivity of CdS.

Applying the boundary conditions, the temperature distribution in the CdSNR becomes:
\begin{equation}
T(x) = \frac{\dot{Q}_{c}}{\kappa HW}x+ T_0.
\label{gradeq}
\end{equation}
It is assumed that the relatively large thermal conductivity of the YLF crystal
($\sim$6 W/m$\cdot$K) will lead to a nearly uniform temperature in the crystal given by:
\begin{equation}
T(L_c) = \frac{\dot{Q}_{c}}{\kappa HW}L_c+T_0.
\end{equation}

The rate of laser energy absorbed per unit volume $Q'''= Q_{abs}/V$ is given by:
\begin{equation}
Q'''=\frac{4\pi n' n''}{\lambda_0 Z_0}(\mathbf{E}\cdot\mathbf{E}^*).
\label{Q'''}
\end{equation}
Here $n = n' - in''$ is the complex refractive index of the medium, $\lambda_0$ is the
wavelength in vacuum, $Z_0$ is the free space impedance ($Z_0$ = 376.73 ohms), and
$\mathbf{E}^*$ is the complex conjugate of the electric field vector within the YLF crystal. Upconverted, anti-Stokes luminescence follows laser absorption, cooling the crystal. We neglect the absorption of the incident laser by the underlying CdS cantilever due to it's small thickness (154 nm) and low absorption coefficient at 1020 nm (6.7x10$^{-13}$ cm$^{-1}$) relative to what has been reported\cite{seletskiy2013precise} for YLF ($\sim$1cm$^{-1}$).

Given that eigenfrequency measurements can only provide a lower bound of the cantilever's temperature, a more direct approach must be used to measure the temperature at the end of the cantilever. Differential luminescence thermometry (DLT)\cite{melgaard2016solid,rahman2017laser} can be used to measure the temperature of the YLF at the end of the cantilever based on using a Boltzmann distribution to analyze emission from different crystal field (Stark) levels. DLT was used to measure a temperature drop of 23.6 K below room temperature ($\Delta T_{max}$) at an irradiance ($I_0$) of 965 kW/cm$^2$ corresponding to an incident power $P_0$ = 40.1 mW and spot radius $w_0$ = 1.15 $\mu$m (see Supplementary Information). Using the measured value of $T(L_c)-T_0$ = 23.6 K, $H$ = 150 nm, $W$ = 2.5 $\mu$m, $L_c$ = 53 $\mu$m,  and $\kappa$ = 20 W/(m K)\cite{moore1969thermal}, we calculate a cooling power of $\dot Q_{c}$ = 3.34 x 10$^{-6}$ W. The resultant temperature gradient along the length of the device is shown in Fig. \ref{F3}b. Based on the temperature gradient, by modeling a spatially varying Young's modulus, the coldest point in the cantilever from eigenfrequency measurements was calculated to be between 26 and 58 K below room temperature (see Supplementary Information). This agrees well with the coldest temperature measured using DLT.

An absorption coefficient and cooling efficiency of 0.61 cm$^{-1}$ and 1.5\%, respectively, have been reported previously for a bulk YLF crystal doped with 10\% Yb-ions.\cite{seletskiy2013precise} Based on this absorption coefficient, a Yb:YLF crystal with a thickness of 6 $\mu$m would generate, to first order, a maximum cooling power of 0.22 $\mu$W when irradiated by a pump laser with a power of 40.1 mW. This power is an order of magnitude smaller than the experimental cooling power reported above. The discrepancy can be explained by two factors related to the symmetric morphology of the YLF microcrystals. First, the size of the YLF microcrystals is within the Mie-regime for light scattering and internal optical fields may be enhanced considerably due to morphology dependent cavity resonances. Figure S11 presents two-dimensional finite-difference time-domain calculations showing that internal optical power within a YLF microcrystal can be twice as large as the incident power due to internal cavity resonances.  Consequently, first-order linear absorption calculations may underestimate the cooling power due to an underestimation of the internal optical power of the pumping laser. Second, a combination of light-scattering and multiple internal reflections of the pump beam within the microcrystal can excite a volume of the crystal that is an order of magnitude greater than the incident spot size. Figure S10 demonstrates that fluorescence is emitted throughout YLF microcrystal, including far from where the excitation laser is focused.  

\section*{Conclusions}
In summary, we demonstrate an approach to refrigerate the temperature of an optomechanical resonator $>$20K below room temperature using solid state laser refrigeration. Thermometry and calibration of the fabricated device were performed using two independent methods - optomechanical eigenfrequencies and differential luminescence thermometry, respectively - which compare well with each other. A modified Euler-Bernoulli model was used to account for the laser trapping forces, and the measured temperatures were validated using heat transfer theory. A maximum drop in temperature of 23.6 K below room temperature was measured near the tip of the cantilever. Among other applications in scanning probe microscopy and exploration of quantum effects at mesoscopic length scales\cite{schwab2005putting}, optical refrigeration of a mechanical resonator could have significant implications for weak force and precision mass sensing applications\cite{braginskii1977measurement,
abramovici1992ligo}, in the development of composite materials for radiation balanced lasers\cite{bowman2005}, and local temperature control in integrated photonic devices\cite{garcia-meca_-chip_2017}. In the future, solid state laser refrigeration may also assist in the cooling of optomechanical devices by enabling the use of higher laser irradiances in the absence of detrimental laser heating.




\begin{methods}
\subsection{Cadmium sulfide nanoribbon synthesis}
The CdSNRs were synthesized using a chemical vapor transport (CVT) method discussed in a previous publication\cite{pant2018optomechanical}. A precursor cadmium sulfide (CdS) powder in an alumina boat which was placed at the center of a quartz tube. The silicon (100) substrates were prepared by dropcasting gold nanocrystals in chloroform. The precursor was heated to 840 $^{\circ}$C. A carrier gas consisting of argon and 5\% hydrogen was used to transport the evaporated species over to a growth substrate placed at the cooler upstream region near the edge of the furnace.

\subsection{Ytterbium doped lithium yttrium fluoride synthesis}
The hydrothermal method used to synthesize single crystals of 10\% ytterbium doped lithium yttrium fluoride (Yb:YLF) nanocrystals was performed following modifications to Roder \textit{et al.}\cite{roder2015laser}. Yttrium chloride (YCl$_3$) hexahydrate and ytterbium chloride hexahydrate (YbCl$_3$) were of 99.999\% and 99.998\% purity, respectively. Lithium fluoride (LiF), lithium hydroxide monohydrate (LiOH.H$_2$O), ammonium bifluoride (NH$_4$HF$_2$), and ethylenediaminetetraacetic acid (EDTA) were analytical grade and used directly in the synthesis without any purification. All chemicals were purchased from Sigma-Aldrich. For the synthesis of Yb:YLF, 0.585 g (2 mmol) of EDTA and 0.168 g (4 mmol) LiOH.H$_2$O were dissolved in 10 mL Millipore DI water and heated to approximately 80 $^\circ$C while stirring. After the EDTA was dissolved, 1.8 mL of 1.0 M YCl$_3$ and 0.2 mL of 1.0 M YbCl$_3$ were added and continually stirred for 1 hour. This mixture is denoted as solution A. Subsequently, 0.105 g (4 mmol) of LiF and 0.34 g (8 mmol) of NH$_4$HF$_2$ were separately dissolved in 5 mL Millipore DI water and heated to approximately 70 $^{\circ}$C while stirring for 1h. This solution is denoted as solution B. After stirring, solution B was then added dropwise into solution A while stirring to form a homogeneous white suspension. After 30 minutes, the combined mixture was then
transferred to a 23 mL Teflon-lined autoclave (Parr 4747 Nickel Autoclave Teflon liner assembly) and heated to 180 $^{\circ}$C for 72 h in an oven (Thermo Scientific Heratherm General Protocol Oven, 65 L). After the autoclave cooled naturally to room temperature, the Yb:YLF particles were sonicated and centrifuged at 4000 rpm with ethanol and Millipore DI water three times. The final white powder was then dried at 60 $^\circ$C for 12 hours followed by calcination at 300 $^\circ$C for 2 hours in a Lindberg blue furnace inside a quartz tube.

\subsection{Device fabrication.}
Using a tungsten dissecting probe (World Precision Instruments) with a sufficiently small tip radius-of-curvature (\textless1 $\mu$m), mounted on to a nano-manipulator
(M\"{a}rzh\"{a}user-Wetzl\"{a}r), the CdSNRs were picked up and placed at the edge of a clean silicon substrate. Yb:YLF crystal was then placed at the free end of the cantilever using the same process.

\subsection{Eigenfrequency measurement.}
The optomechanical thermometry setup consists of a 1020 nm diode laser (QPhotonics) focused, using a 50x long working distance objective, onto the sample placed inside a cryostat (Janis ST500) modified by drilling an axial hole through the sample stage. The forward scattered light was collected through the axial hole and was focused onto an Avalanche photo-diode (APD, Thorlabs APD430A). The time domain voltage signal from the APD was then Fourier-transformed to obtain the thermomechancial noise spectrum with characteristic peaks from the fundamental eigenfrequency modes of the cantilever. For temperature calibration, the thermomechanical noise spectrum was
recorded by varying the cryostat temperature and fitting the resulting peaks using a standard Lorentzian. Each data point represents the average of six measurements and the error bars represent the standard deviation.

\end{methods}
\begin{addendum}
 \item A.P., X.X., and P.J.P. gratefully acknowledge financial support from the MURI:MARBLe project under the auspices of the Air Force Office of Scientific Research (Award No. FA9550-16-1-0362). Sample characterization was conducted at the University of Washington Molecular Analysis Facility, which is supported
in part by the National Science Foundation (Grant No. ECC-1542101), the University of Washington, the Molecular Engineering \& Sciences Institute, the Clean Energy Institute,
and the National Institutes of Health.
 
 \item[Competing Interests] The authors declare that they have no
competing financial interests.
 \item[Correspondence] Correspondence and requests for materials
should be addressed to Peter J. Pauzauskie.~(email: peterpz@uw.edu).
\end{addendum}


\end{document}